\documentclass[pdftex,iop,apjl]{emulateapj}

\shorttitle{Comet Dust and Gas at Mars}

\shortauthors{Kelley et al.}

\submitted{Accepted for publication in Astrophysical Journal Letters,
  5 Aug 2014}

\usepackage{amsmath}

\begin{document}

\title{A Study of Dust and Gas at Mars from Comet C/2013 A1 (Siding
  Spring)}

\author{Michael S. P. Kelley, Tony L. Farnham, and Dennis Bodewits}
\affil{Department of Astronomy, University of Maryland, College Park,
  MD 20742-2421, USA}
\email{msk@astro.umd.edu}

\author{Pasquale Tricarico}
\affil{Planetary Science Institute, 1700 E. Ft. Lowell \#106, Tucson,
  AZ 85719, USA}

\author{Davide Farnocchia}
\affil{Jet Propulsion Laboratory, California Institute of Technology,
  4800 Oak Grove Drive, Pasadena, CA 91109, USA}

\begin{abstract}
  Although the nucleus of comet C/2013 A1 (Siding Spring) will safely
  pass Mars in October 2014, the dust in the coma and tail will more
  closely approach the planet.  Using a dynamical model of comet dust,
  we estimate the impact fluence.  Based on our nominal model no
  impacts are expected at Mars.  Relaxing our nominal model's
  parameters, the fluence is no greater than
  $\sim10^{-7}$~grains\,m$^{-2}$ for grain radii larger than
  10~\micron.  Mars orbiting spacecraft are unlikely to be impacted by
  large dust grains, but Mars may receive as many as $\sim10^7$
  grains, or $\sim100$~kg of total dust.  We also estimate the flux of
  impacting gas molecules commonly observed in comet comae.

\end{abstract}


\section{INTRODUCTION}

Comet C/2013 A1 (Siding Spring) will pass Mars with a close approach
distance of $1.35\pm0.05\times10^5$~km, and a relative speed of
55.96~km\,s$^{-1}$ on 2014 Oct 19 at 18:29$\pm$:03 UTC
\citep[3-$\sigma$ uncertainties;][]{farnocchia14}.  The nucleus will
miss the planet, its moons, and orbiting spacecraft.  However, given
the right combination of ejection velocity, ejection time, and
response to radiation pressure, dust grains from the comet can reach
the planet.  \citet{farnocchia14} predict that Mars will miss the
comet's orbit by $2.7\times10^4$~km at 20:10 UTC.  This second close
approach potentially reduces the energy required to place dust grains
on impacting orbits.  We present models of the dust and gas based on
the summary of the comet's activity by Farnham et al. (in preparation)
and estimate the impact hazard for Mars and its satellites as well as
the comet gas flux at Mars.

\section{SIMULATIONS}
\subsection{Dust Dynamics}
To assess the impact hazard, we generated two simulations of $10^9$
particles each, picked from broad parameter ranges.  These raw
simulations act as guides to determine which combinations of size,
ejection speed, and ejection time may result in impacts.  Next, we
define more limited parameter sets that are carefully chosen to match
known parameters of the comet.  We select and weight particles from
the raw simulations that match those sets, and use them to estimate
the fluence at Mars.  Below we describe our dynamical model, the raw
simulations, and four parameter sets used to estimate the impact
hazard.

The circumstances of the encounter are simulated with the dynamical
model of \citet{kelley06-phd}.  For this study we use the JPL
ephemeris solution \#46 \citep{farnocchia14}.  In order to reduce the
required computational time, we modified the model to use the two-body
(Keplerian) propagation functions from NASA's Navigation and Ancillary
Information Facility SPICE toolkit.  Dust grains are parameterized by
$\beta$, the ratio of the force from solar radiation pressure to the
force from solar gravity: $\beta=0.57 Q_{pr}/\rho a$, where $Q_{pr}$
is the radiation pressure efficiency, $\rho$ is the grain density in
units of g\,cm$^{-3}$, and $a$ is the grain radius in units of
\micron{} \citep{burns79}.  In the Keplerian solution, the
gravitational force from the Sun is reduced by the factor $(1 -
\beta)$.

The magnitude of the error introduced by neglecting planetary
perturbations can be estimated by comparing zero-ejection velocity
syndynes \citep[lines of constant $\beta$ with variable ejection
times;][]{finson68-paper1} generated using the Keplerian solution to
those generated using the original code.  The distances between the
syndynes define the error.  For grains ejected up to 4 years before
the closest approach, the error is at most 300~km for dust found
within $10^6$~km from the nucleus.  We also considered whether the
gravitational pull of Mars is significant.  Ignoring the atmosphere,
particles grazing the surface are displaced $<100$~km at closest
approach, and the cross-section enhancement factor from gravitational
focusing by Mars is 1.008 \citep{jones07}.  The Keplerian solution is
sufficient for our purposes.

Simulation~1 contains $10^9$ particles selected from the
following parameters, based on observations of the comet with a
generous conservative margin: ages range uniformly from 0 to 4~yr (out
to $r_h=13$~AU); expansion speeds range uniformly from 0 to
$v_{ref}\,(a/1\mbox{~mm})^{-0.5}\,(r_h/5\mbox{~AU})^{-1}$, where
$v_{ref}=1.9$~m\,s$^{-1}$ is the expansion speed of 1~mm grains
ejected at 5~AU from the Sun; ejection velocities are radial and
isotropically distributed around the nucleus; and, radii are selected
from a distribution uniform in log-space
($\mathrm{d}n/\mathrm{d}\log{a}\propto1$) ranging from 10 to
$10^4$~\micron.  The logarithmic distribution ensures our final
results will have a statistically uniform representation of each size
decade.  For the conversion from radius to $\beta$ we assume a grain
density of 1~g\,cm$^{-3}$, and $Q_{pr}=1$.  J.-Y.~Li et al. (in
preparation) imaged Siding Spring at 4.6, 3.8, and 3.3~AU from the Sun
with the \textit{Hubble Space Telescope} WFC3 instrument.  The high
spatial resolution of the images (40~mas per pixel, corresponding to
$\geq100$~km per pixel) resolve the inner coma, and allow
investigations of the dust grain expansion velocities.  Farnham et
al. (in preparation) analyzed those images and found that the dust
that comprises the bulk of the coma and tail has speeds best matched
by $v_{ref}\,(a/1\mbox{~mm})^{-0.6}\,(r_h/5\mbox{~AU})^{-1.5}$, for
$v_{ref}=0.42$~m\,s$^{-1}$.  Grains with these speeds are a subset of
simulation~1.

For an alternative scenario we consider the analysis of the
\textit{Hubble}{} images by Li et al..  Using the distance from the
nucleus to the outer edge of the coma in the sunward direction, they
find that the heliocentric distance dependence of the resulting speeds
follow $v_{ej}\approx800\,r_h^{-2}$~m\,s$^{-1}${} for
$r_h=3.3-4.6$~AU, assuming $\beta=1$ grains
($v_{ref}\approx0.76$~m\,s$^{-1}$).  Speeds based on this relationship
for $r_h<2.0$~AU exceed those in simulation~1.  Therefore, we ran a
second $10^9$ particle simulation (simulation 2) with speeds picked
uniformly from 0 to
$v_{ref}\,(a/1\mbox{~mm})^{-0.5}\,(r_h/5\mbox{~AU})^{-2}$ for
$v_{ref}=1.9$~m\,s$^{-1}$.  However, we note that these speeds,
derived from some of the fastest moving grains in the coma, are much
higher than the Farnham et al. results, derived from the coma and tail
morphologies.  Analyses based on the Li et al.\ speeds will serve as
upper-limit cases for fast moving grains not accounted for in the
Farnham et al. approach.

\begin{figure*}
  \centering
  \includegraphics[width=\textwidth]{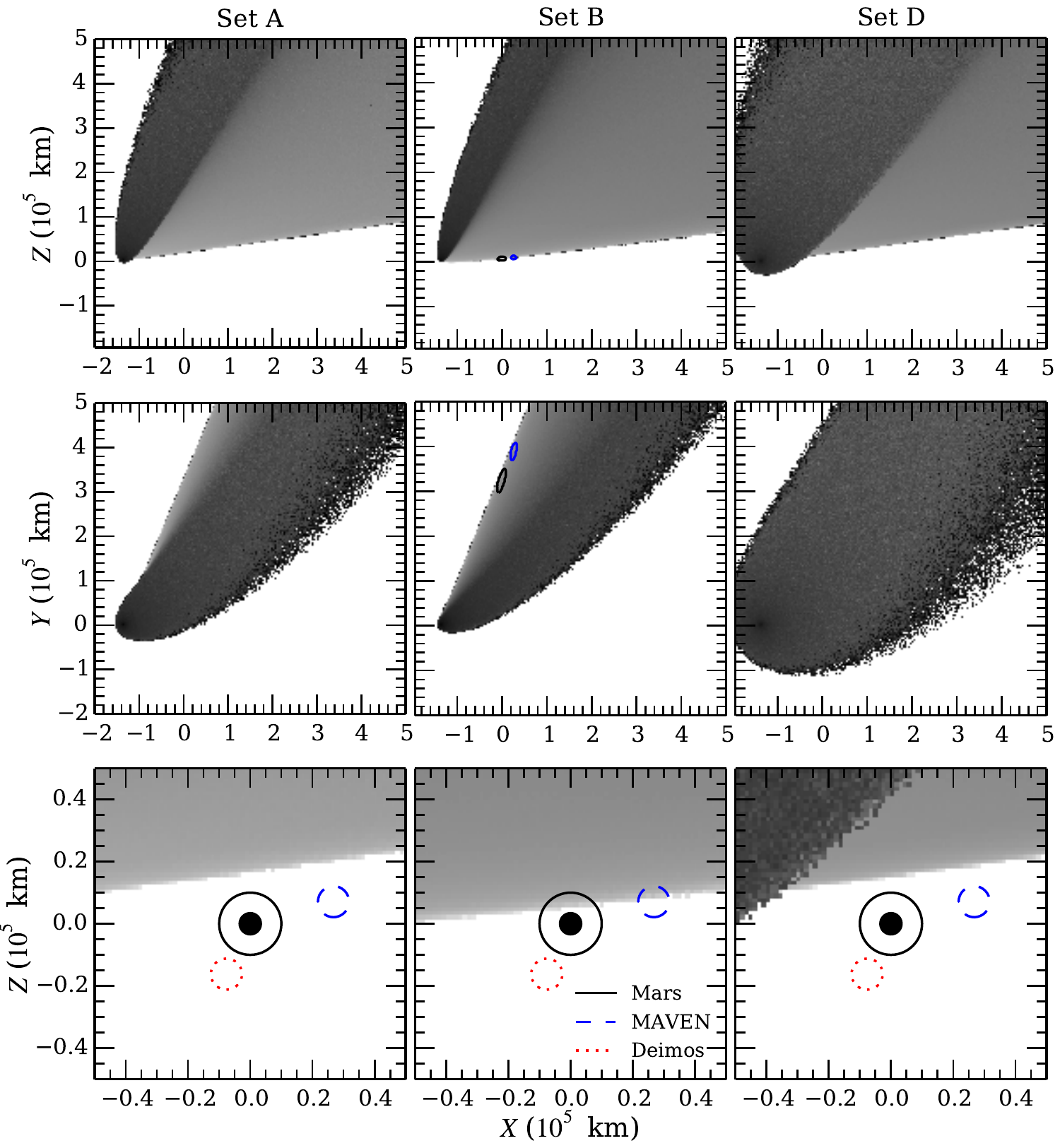}
  \caption{\small Distribution of simulated comet dust rotated into
    our closest approach reference frame.  Mars is located at $(0, 0,
    0)$.  Column labels indicate the parameter set used from
    Table~\ref{tab:parameters}.  The logarithmic gray scale is
    illustrative (black indicates more particles), weighted to reflect
    a size distribution of $\mathrm{d}n/\mathrm{d}a\propto{}a^{-4}$,
    but without any other scaling.  (Top and center rows) Contours
    encircle the population of grains within our Mars, Deimos, and
    \textit{MAVEN} regions of interest.  Sets A and D have no
    contours, indicating no hazardous grains.  (Bottom row) The
    closest 50,000~km to Mars.  The surface of Mars is indicated with
    a solid black disk.  The Mars, Deimos, and \textit{MAVEN} regions
    of interest are indicated with circles.  The orbit of Phobos is
    contained within the Mars region of interest. \label{fig:xyz}}
\end{figure*}

In order to transform the simulations into an impact hazard at Mars,
we first rotate the position vectors from the ecliptic J2000
coordinate frame into a reference frame defined at closest approach:
the x-axis is given by the comet-Mars position vector, the y-axis by
the comet-Mars velocity vector, and the z-axis by the right-hand rule.
Simulations 1 and 2 projected into this reference frame is shown in
Fig.~\ref{fig:xyz}.  The relative timing of grains arriving at Mars is
based on their y-axis position; $10^5$~km corresponds to a time
difference of 29.8~min.  We then weight each particle to remove the
bias introduced by our raw simulation's grain size distribution, and
to provide a real estimate of the grain's frequency of occurrence in
the comet coma.  The particle weights are based on the parameter sets
in Table~\ref{tab:parameters}.

Set A is the nominal case, directly based on the results of Farnham et
al. (in preparation).  Sets B, C, and D are variations chosen to
provide upper-limit estimates on the impacting dust.  First, consider
that the tail analysis of Farnham et al. best describes the smallest
grains ($a\lesssim 10$~\micron) and closest heliocentric distances
(3~AU $\leq r_h \lesssim$ 5~AU), and is extrapolated to larger $a$ and
$r_h$.  Therefore, as an alternative scenario, we have defined set B
using a radius and heliocentric distance dependence in closer
agreement to theoretical predictions:
$v_{ej}\propto{}a^{-0.5}\,r_h^{-1}$ \citep[cf.][]{whipple51-meteors,
  crifo97-paper1}.  The speeds follow
$v_{ej}=9.4\,(\beta/0.1)^{0.5}\,(r_h/5\mbox{~AU})^{-1}$~m\,s$^{-1}$,
corresponding to $v_{ref}=0.71$ for $\beta<0.1$ ($a>6$~\micron).

Set C has the same parameters as set B, but the dust is shifted closer
to Mars according to the ephemeris's 3-$\sigma$ error ellipse
\citep{farnocchia14}.  In our closest approach reference frame
described above, the dust is displaced by $\Delta{}X=4471$~km,
$\Delta{}Z=-1722$~km.

Set D is based on our nominal case, but uses simulation 2 and the
ejection speeds found by Li et al.\ (in preparation).  As described
above, these speeds are representative of the fastest moving grains,
and not the whole coma.

\begin{deluxetable*}{lcccc}
  \tabletypesize{\footnotesize}
  \tablecaption{Model parameters and results.\label{tab:parameters}}
  \tablewidth{0pt}
  \tablehead{
    \colhead{Parameter}
    & \colhead{Set A}
    & \colhead{Set B}
    & \colhead{Set C}
    & \colhead{Set D}
  }

  \startdata
  Gas production rate at 3.1~AU, $Q_g$ (kg\,s$^{-1}$)
  & 25.7
  & \nodata
  & \nodata 
  & \nodata \\

  $Q_g$ heliocentric dependence for $r_h \leq 6.1$~AU
  & $r_h^0$
  & \nodata
  & \nodata
  & \nodata \\

  $Q_g$ heliocentric dependence for $r_h > 6.1$~AU
  & $r_h^{-7}$
  & \nodata
  & \nodata
  & \nodata \\

  Dust-to-gas mass ratio
  & 1.0
  & \nodata
  & \nodata
  & \nodata \\

  Reference expansion speed,\tablenotemark{a} $v_{ref}$ (m\,s$^{-1}$)
  & $0.42\pm0.02$
  & $0.71\pm0.07$
  & $0.71\pm0.07$
  & $0.76\pm0.04$ \\

  Expansion speed dependence on $a$
  & $a^{-0.6}$
  & $a^{-0.5}$
  & $a^{-0.5}$
  & $a^{-0.5}$ \\

  Expansion speed dependence on $r_h$
  & $r_h^{-1.5}$
  & $r_h^{-1.0}$
  & $r_h^{-1.0}$
  & $r_h^{-2.0}$ \\

  Minimum grain radius (\micron)
  & 0.1
  & \nodata
  & \nodata
  & \nodata \\

  Maximum grain radius (\micron)
  & $10^4$
  & \nodata
  & \nodata
  & \nodata \\

  Grain size distribution, $\mathrm{d}n/\mathrm{d}{}a$
  & $a^{-4}$
  & \nodata 
  & \nodata 
  & \nodata \\

  Comet-Mars closest-approach distance (km)
  & $1.35\times10^5$
  & \nodata
  & $1.30\times10^5$
  & \nodata \\

  \hline
  Raw (unweighted) number of particles\tablenotemark{b}
  & $5.9\times10^5$
  & \nodata
  & $7.7\times10^5$
  & $8.4\times10^3$ \\

  Raw (unweighted) number of impacting grains\tablenotemark{b}
  & 0
  & $4.5\times10^3$
  & $1.5\times10^4$
  & 0 \\

  Total fluence\tablenotemark{b,c} ~~($10^{-7}$ grains\,m$^{-2}$)
  & 0
  & $1.14\pm0.02$
  & $3.91\pm0.03$
  & 0 \\

  Total fluence\tablenotemark{b,c} ~~($10^{-12}$ kg\,m$^{-2}$)
  & 0
  & $3.27\pm0.05$
  & $11.7\pm0.1$
  & 0 \\

  Time of impacts\tablenotemark{b} (UTC)
  & NA
  & 19:57--20:17
  & 19:52--20:11
  & NA

  \enddata

  \tablecomments{Ellipses indicate the parameter is unchanged from set
    A.}

  \tablenotetext{a}{Expansion speed of 1~mm-radius grains at 5~AU.}

  \tablenotetext{b}{After removing particles that do not meet the
    ejection speed criteria.  Fluence and times are given at Mars.
    Times were calculated using 5-min bins, where ``NA'' indicates the
    time cannot be computed.}

  \tablenotetext{c}{Fluence uncertainties are based on Poisson
    statistics, and do not represent uncertainties inherent to the
    model or inputs.}

\end{deluxetable*}

\subsection{Gas}
The future activity of comet Siding Spring is challenging to predict,
but must be based on our present set of observations.  The gas
production rates, $Q(\mbox{CO$_2$})=(3.52\pm0.03)\times
10^{26}$~molecules\,s$^{-1}${} at $r_h=3.1$~AU, and $Q(\mbox{H$_2$O})=(1.7\pm
1)\times10^{27}$~molecules\,s$^{-1}${} at $r_h=2.5$~AU, were measured by Farnham
et al. (in preparation) and \citet{bodewits14-cbet}, respectively,
based on photometric imaging of the comet with the \textit{Spitzer
  Space Telescope} and the \textit{Swift} satellite.  We make two
predictions of the gas production rate at Mars, based on these
measurements of CO$_2${} and H$_2$O.

First, we propagate the measured CO$_2${} production rate from 3.1~AU to
1.4~AU, then compute the water production rate using an assumed
CO$_2$-to-H$_2$O{} mixing ratio.  The mean mixing ratio, based on
Table~1 of \citet{ahearn12-origins} is
$Q(\mbox{CO$_2$})/Q(\mbox{H$_2$O})=0.14$.  Based on pre-perihelion
observations of 14 dynamically new comets, \citet{whipple78} found
that, on average, their lightcurves grow as $r_h^{-2.44}$, suggesting
activity grows as $\sim r_h^{-0.44}$.  We adopt $Q\propto r_h^0$ and
$\propto r_h^{-1}$ to derive a range of possible values at 1.4~AU:
$Q(\mbox{CO$_2$})=3-8\times10^{26}$~molecules\,s$^{-1}${} and
$Q(\mbox{H$_2$O})=2-6\times10^{27}$~molecules\,s$^{-1}${}.  Repeating the
exercise, but instead starting with the water measurement at 2.5~AU,
we find $Q(\mbox{CO$_2$})=2-4\times10^{26}$~molecules\,s$^{-1}${} and
$Q(\mbox{H$_2$O})=2-3\times10^{27}$~molecules\,s$^{-1}${}.  Altogether, we adopt
$Q(\mbox{CO$_2$})=(5\pm2)\times10^{26}$~molecules\,s$^{-1}${} and
$Q(\mbox{H$_2$O})=(4\pm1)\times10^{27}$~molecules\,s$^{-1}${}.

To estimate the gas fluence and peak volume density at the top of the
Martian atmosphere, we modeled the coma with a modified
three-generation Haser model \citep{festou81, combi04,
  bodewits11-lulin}, considering H$_2$O, CO$_2$, CO, and CN.  The
relative abundances of CO and CN are based on the mean CO-to-H$_2$O{}
mixing ratio (0.07) from \citet{ahearn12-origins}, and the ``typical''
composition of dynamically new comets from \citet{ahearn95}.  We also
include OH, H, and O as photodissociation products of H$_2$O.  Our
results are given in Table~\ref{tab:gas}.

\begin{deluxetable*}{lcccc}
  \tabletypesize{\footnotesize} 
  \tablecaption{Total comet gas column densities, and peak fluxes at
    Mars.\label{tab:gas}}
  \tablewidth{0pt}
  \tablehead{
    \colhead{Gas}
    & \colhead{Production Rate}
    & \colhead{Relative Abundance}
    & \colhead{$\Sigma$}
    & \colhead{$F_{peak}$} \\
    & \colhead{($10^{27}$ s$^{-1}$)}
    & 
    & \colhead{($10^{15}$ m$^{-2}$)}
    & \colhead{($10^{11}$ m$^{-2}$\,s$^{-1}$)}
  }

  \startdata

  H$_2$O & 4.0     & 100  & 1.9    & 4.4   \\
  OH     & 0       & 0    & 2.8    & 4.4   \\
  O      & 0       & 0    & 2.0    & 1.7   \\
  H      & 0       & 0    & 0.94   & 0.88  \\
  CO$_2$ & 0.50    &  13  & 8.2    & 14.   \\
  CO     & 0.28    &   7  & 0.38   & 0.70  \\
  CN     & 0.012   & 0.3  & 0.0098 & 0.019 
  \enddata

  \tablecomments{OH is a photodissociation product of H$_2$O.  Only
    photodissociation of H$_2$O{} and OH are considered for the H and
    O abundances.}

\end{deluxetable*}

\section{RESULTS}
\subsection{Raw Simulations}
\begin{figure}
  \centering
  \includegraphics[width=\columnwidth]{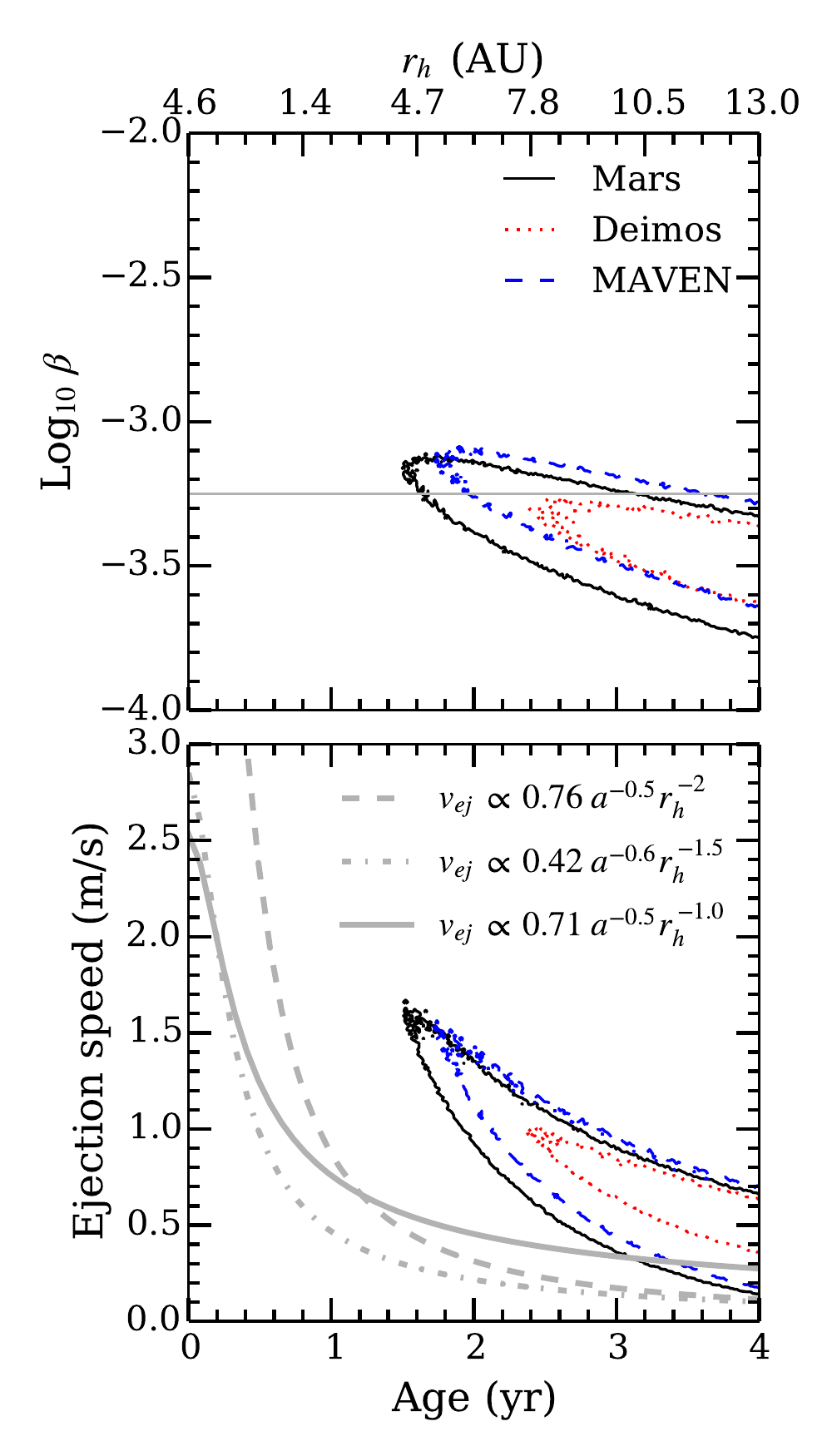}
  \caption{The $\beta$, age, and ejection speed of particles from
    simulation~1 found within our Mars, Deimos, and \textit{MAVEN}
    regions.  The parameter ranges enclosed within the contours yield
    potential impact hazards.  (Top) A horizontal line indicates the
    $\beta$ value of 1-mm radius grains.  (Bottom) Additional lines
    show the ejection speeds of 1-mm-sized grains, based on parameter
    sets A, B, and D. \label{fig:beta-age-speed}}
\end{figure}

The raw simulations provide a guide to understanding what combination
of grain size, ejection velocity, and age results in an impact hazard
at Mars.  We define impacts at Mars as any particle found within a
distance of 10,000~km from the center of the planet.  This distance
includes the orbit of Phobos (9400~km semi-major axis), and the
apoapsis of the \textit{MAVEN} spacecraft's nominal science orbit
(6000~km; provided by NASA JPL).  We also investigate a 5000~km region
centered on the position of Deimos (23,000~km semi-major axis) at the
time of closest approach.  A third region of interest is based on the
48,000~km apoapsis of \textit{MAVEN}'s orbit, assuming a 5-week delay
in science operations.  Given these criteria, the grain parameters
that yield impacts are presented in Fig.~\ref{fig:beta-age-speed}.
The only particles in simulation~1 that reach the Martian system are
those ejected with speeds of a few meters per second, have radii of
0.7--3.6~mm, and are ejected at least 1.5~yr prior to the encounter.

\subsection{Impact Hazard}
For each parameter set, the impact hazards are computed by taking the
set of particles found in each region of interest, removing those
speeds outside the set's range, and weighting remaining particles
according to the set's total production rate, grain size distribution,
etc.  Table~\ref{tab:parameters} lists, for Mars, the raw number of
impacting grains (i.e., before particle weighting), the impact fluence
(i.e., after particle weighting), and the start and stop times of the
hazard.  No impacts are expected based on our nominal model, nor the
$v_{ej}\propto~r_h^{-2}$ model (sets A and D) because the ejection
speeds of dust grains at $r_h>3$~AU are too low to place particles
within the vicinity of Mars (Fig.~\ref{fig:beta-age-speed}).  In
addition, displacing the comet dust in these two models according to
the ephemeris 3-$\sigma$ uncertainty ellipse does not result in any
impacts.

In order to attain impacting particle trajectories, higher ejection
speeds are needed at $r_h>3$~AU.  This requirement is accomplished
with the size-$r_h$-speed relationship of parameter set B, resulting
in a total fluence of $1\times10^{-7}$~grains\,m$^{-2}$.  Based on the
comet's current predicted closest approach time, the grains arrive
between 19:57 and 20:17 UTC (time at Mars), nearly centered on the
epoch at which Mars crosses the comet's orbital plane
\citep{farnocchia14}.  Note that non-gravitational forces frequently
act upon comet nuclei, but we can neglect these effects because the
dust is ejected at heliocentric distances where activity and resulting
non-gravitational forces were low \citep[see Fig.~5
of][]{farnocchia14}.  However, the current nucleus ephemeris
uncertainty is still valid for the dust.  Displacing the dust in
parameter set B closer to Mars (set C) shifts the arrival time 5~min
earlier and increases the fluence by a factor of 4.

\begin{figure*}
  \centering
  \includegraphics[width=6in]{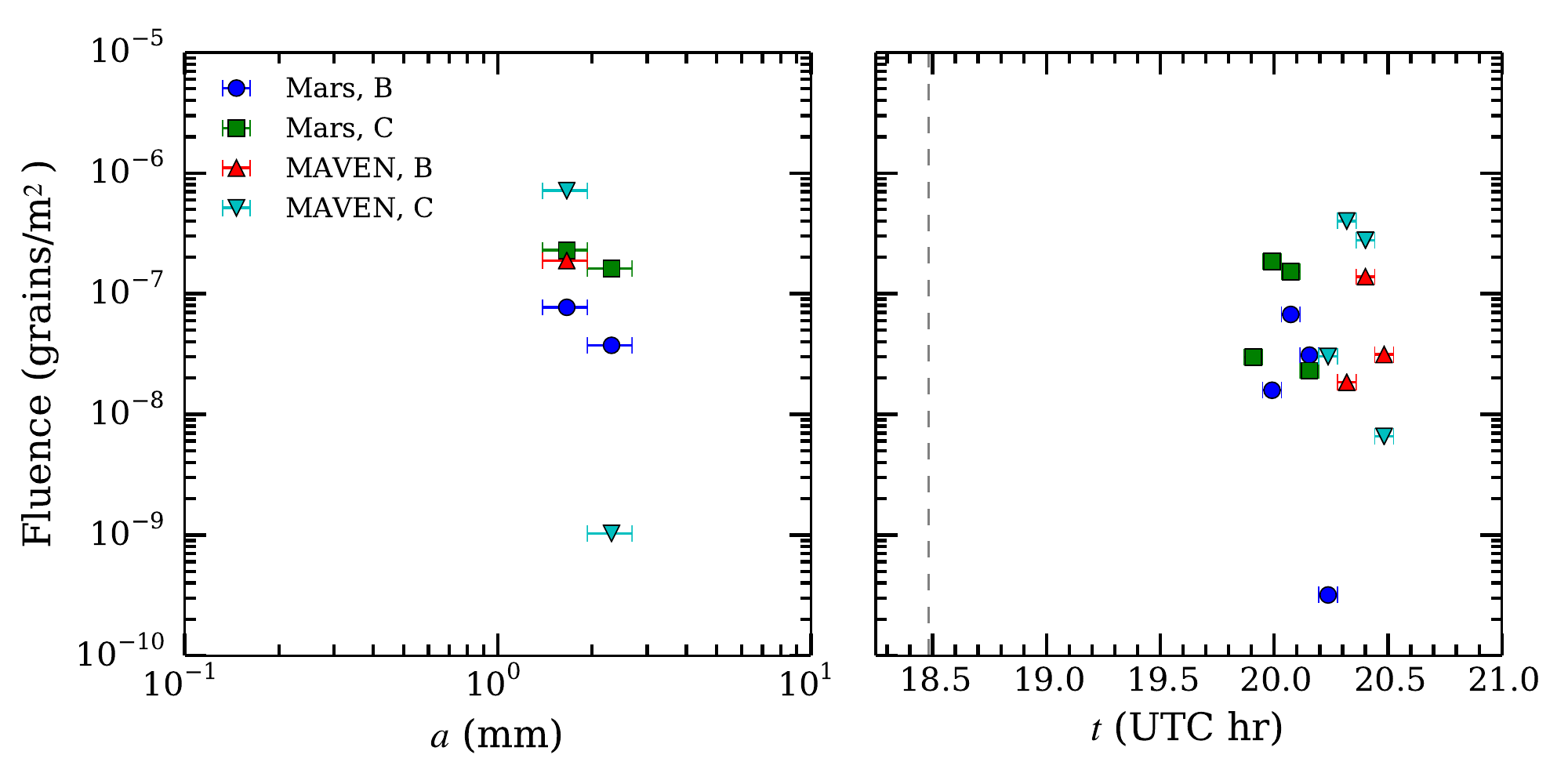}
  \caption{(Left) Dust grain fluence in our regions of interest versus
    grain radius computed for parameter sets B and C (sets A and D
    have no impacts).  Error bars indicate the bin width over which
    the flux was summed. (Right) Fluence versus time, in 5-min
    bins. \label{fig:fluence}}
\end{figure*}

The fluence results are summarized for Mars and \textit{MAVEN} in
Fig.~\ref{fig:fluence}.  The grains are limited to 1 to 3~mm in
radius, due to the combined effects of radiation pressure, grain age,
and ejection speed.  The timing of the hazard at \textit{MAVEN}'s distant
apoapsis is 25~min later than the hazard at Mars, and the fluences are
a factor of 2 higher.  In all simulations, the Deimos region of
interest remained dust free.

Gas outflow and nucleus gravity limits the size of dust grains that
can be placed into heliocentric orbit.  \citet{meech04-activity}
present the critical dust radius, $a_{crit}$, that can be lifted off a
spherical nucleus by estimating the drag force on a spherical grain,
and integrating the resulting equation of motion,
\begin{equation}
  a_{crit} = \frac{9 \mu m_{\textrm{H}} Q v_{th}}{64 \pi^2 \rho_g \rho_n
    R_n^3 G},
\end{equation}
where $\mu$ is the atmoic weight of the driving gas, $m_{\textrm{H}}$
is the mass of hydrogen, $Q$ is the gas production rate
(molecules\,s$^{-1}$), $v_{th}$ is the expansion speed of the gas,
$\rho_g$ and $\rho_n$ are the grain and nucleus densities, $R_n$ is
the nucleus radius, and $G$ is the gravitation constant.  Let
$v_{th}=0.6~(r_h/3.8\textrm{~AU})^{-0.5}$~km\,s$^{-1}$ to be
consistent with Farnham et al.\ (in preparation), and let the nucleus
density be 0.3~g\,cm$^{-3}$.  The radius of the nucleus has not been
measured, but \citet{bodewits14-cbet} estimate $R>0.34$~km based on
water projection rates from \textit{Swift} photometry.  Therefore we
consider two values: 0.5 and 2.0~km.  For our adopted parameters,
CO$_2${} can lift grains larger than 100~\micron{} at heliocentric
distances inside of 14 and 8~AU for $R_n=0.5$ and 2.0~km,
respectively.  However, our parameter set ejection speeds only place
particles on impacting trajectories for $r_h>11$~AU
(Fig.~\ref{fig:beta-age-speed}).  Thus, given this simplistic model,
an impact hazard may not be expected if the nucleus radius is
$\sim2$~km or larger.

Overall, we do not expect any impacts on Mars-orbiting spacecraft.
Based on the cross-sectional area of Mars and our 10,000~km average
fluence, the planet may receive up to $\sim10^7$ grain impacts from 1-
to 3-mm-radius grains, totaling $\sim100$~kg, based on our models.  At
most, a few impacts may be expected on Phobos ($\lesssim100$), and no
impacts at Deimos.

\subsection{Comparison with Other Results}
\citet{moorhead14} and \citet{vaubaillon14} predict a significantly
larger impact hazard, with fluences of 0.1~grains\,m$^{-2}$ and
larger.  Their large fluences appear to be primarily due to their
choice of ejection speeds.  \citet{moorhead14} use $v_{ref}=11$~m\,s$^{-1}${}
(A.~Moorhead, private communication), and \citet{vaubaillon14} use
$v_{ref}$ up to 20~m\,s$^{-1}${} (J.~Vaubaillon, private communication).
Unfortunately, few observations that could constrain the dust
expansion speeds, if any, were available to these investigators.

In contrast, \citet{ye14} and \citet{tricarico14} constrain their
fluence estimates using observations of the comet.
\citet{tricarico14} base their methods on the Farnham et al. (in
preparation) and Li et al. (in preparation) data, the same as our
investigation, but with slightly different interpretations.  They both
predict little to no risk of impacts for the Mars-orbiting spacecraft,
despite their independent approaches.  In particular, \citet{ye14}
used observations of Siding Spring and the similarly bright comet
C/2012 S1 (ISON) to derive the dependence of ejection speed on size
and $r_h$:
\begin{align}
  v_{ej}=\,& 1.0\textrm{~m\,s$^{-1}$} (a/5\textrm{~mm})^{-0.5}
  (r_h/1\textrm{~AU})^{-1} \nonumber\\
  &(\rho/1\textrm{~g\,cm$^{-3}$})^{-0.5} (R_n/0.5\textrm{~km})^{0.5}
\end{align}
best matched their data, assuming a 2.5-km-radius nucleus and
0.3-g\,cm$^{-3}${} dust, resulting in no impacts for $a>0.1$~mm, and a
fluence of $2.6\times10^{-6}$~grains\,m$^{-2}$ for grains down to
10~\micron{} in radius.  In our parameterization, these speeds
correspond to $v_{ref}=1.0$~m\,s$^{-1}${}, higher than our set B.  Using
their speeds and our production rate history, we find the same
fluence, $2\times10^{-6}$~grains\,m$^{-2}$, but all impacts are
millimeter sized.

Finally, we compare our results to the natural background of
meteoroids estimated for the Mars Reconnaissance Orbiter.  Over a
5-year period, the total fluence for meteoroids with $a>1$~mm is
0.0021~grains\,m$^{-2}$, and for $a>0.1$~mm, it is
3.1~grains\,m$^{-2}$ \citep{mro-doc}.  Our predicted fluences from
comet Siding Spring are orders of magnitude smaller than these values,
and we conclude the comet poses little additional hazard to the
spacecraft.

\subsection{Comet Gases at Mars}
The neutral coma gases will enter the atmosphere with a relative
velocity of 56~km\,s$^{-1}$.  The kinetic energies of these molecules
greatly exceeds their dissociation energies, e.g., for H$_2$O, the
kinetic energy is 293 eV, and the dissociation energy (H$_2$O
$\rightarrow$ OH + H) is 5~eV \citep{darwent70}.  The upper atmosphere
of Mars consists of CO$_2$, with few percent contributions from N$_2$
and Ar, and trace amounts of several other species
\citep{krasnopolsky02}.  The comet gases will collide with the Mars
gases, and quickly dissociate into atoms, erasing any molecular trace
of the comet in the Martian atmosphere.  The peak total particle and
kinetic energy fluxes in Table~\ref{tab:gas} are
$2.6\times10^{12}$~m$^{-2}$\,s$^{-1}$ and
$2.0\times10^{-4}$~W\,m$^{-2}$, respectively, comparable to the solar
wind at Mars: $1.8\times10^{12}$~m$^{-2}$\,s$^{-1}$ and
$2.3\times10^{-4}$~W\,m$^{-2}$ for 4.4~protons\,cm$^{-3}$ moving at
400~km\,s$^{-1}$.  Effects in the Martian atmosphere caused by the
impacting comet gases are discussed by \citet{yelle14}.

\subsection{Summary}
Based on the observations of Farnham et al. (in preparation) and Li et
al. (in preparation), we simulated the coma and tail of comet Siding
Spring for grains with radii between 10~\micron{} and 1~cm
($\rho=1.0$~g\,cm$^{-3}$), and particle ages out to 4 years (13 AU)
before closest approach to Mars.  We predict no dust impacts at Mars
from the close flyby of the comet in October 2014.  Variations of our
nominal comet model suggest a total fluence of
$\lesssim10^{-7}$~grains\,m$^{-2}$ is possible, with radii ranging
from 1 to 3~mm, and encounter times between 19:52 and 20:17 UTC (time
at Mars).  Mars orbiting spacecraft are unlikely to be impacted by any
large dust grains, but Mars may receive as many as $\sim10^7$ grains
($\sim100$~kg).  Following \citet{vaubaillon14}, the meteor shower at
Mars is an Earth-equivalent zenith hourly rate $\lesssim600$~h$^{-1}$
\citep[assuming a human perception correction factor of 3 given our
  meteor size range;][]{koschack90-paper1}. The gas coma will reach
the upper atmosphere of Mars with peak fluxes of order
$10^{12}$~molecules\,m$^{-2}$\,s$^{-1}$, and the molecules will be
quickly dissociated, due to the high impact speeds.

\acknowledgments

This research was supported by a contract to the University of
Maryland by the NASA JPL Mars Critical Data Products Program.

The work of D.~Farnocchia was conducted at the Jet Propulsion
Laboratory, California Institute of Technology under a contract with
NASA.

Simulations were performed on the YORP cluster administered by the
Center for Theory and Computation, part of the Department of Astronomy
at the University of Maryland.

This research made use of Astropy, a community-developed core Python
package for Astronomy \citep{astropy13}.

\end{document}